\documentclass[useAMS,usenatbib,twocolumn]{mn2e}
\usepackage{amssymb}
\usepackage{amsmath} 
\usepackage{graphicx}
\usepackage{lscape}
\usepackage{longtable}
\usepackage{array}
\usepackage{color}
\title[Further evidence for a time-dependent IMF in ETGs]
{Further evidence for a time-dependent initial mass function in 
massive early-type galaxies}

\author[I.~Ferreras et al.]
{Ignacio Ferreras$^{1,2}$\thanks{E-mail: i.ferreras@ucl.ac.uk}, 
Carsten Weidner$^{3,4}$, Alexandre Vazdekis$^{3,4}$, Francesco La Barbera$^5$\\
$^1$Mullard Space Science Laboratory, University College London,
Holmbury St Mary, Dorking, Surrey RH5 6NT\\
$^2$Severo Ochoa visitor, Instituto de Astrof\'\i sica de Canarias\\
$^3$Instituto de Astrof{\'i}sica de Canarias, Calle V{\'i}a L{\'a}ctea s/n, 
E38205, La Laguna, Tenerife, Spain\\
$^4$Dept. Astrof{\'i}sica, Universidad de La Laguna (ULL), E-38206 La Laguna, 
Tenerife, Spain\\
$^5$INAF--Osservatorio Astronomico di Capodimonte, I-80131 Napoli, Italy
}

\begin{document}
\bibliographystyle{aa}
\date{Accepted 2014 January 06. Received 2014 January 05; in original
  form 2014 September 19}

\pagerange{\pageref{firstpage}--\pageref{lastpage}} \pubyear{2015}

\maketitle

\label{firstpage}

\begin{abstract}
Spectroscopic analyses of gravity-sensitive line strengths give
growing evidence towards an excess of low-mass stars in massive
early-type galaxies (ETGs). Such a scenario requires a bottom-heavy
initial mass function (IMF). However, strong constraints can be
imposed if we take into account galactic chemical enrichment. We
extend the analysis of Weidner et al. and consider the functional form
of bottom-heavy IMFs used in recent works, where the high-mass end
slope is kept fixed to the Salpeter value, and a free parameter is
introduced to describe the slope at stellar masses below some pivot
mass scale (M$<$M$_{\rm P}=0.5$M$_\odot$). We find that no such
time-independent parameterisation is capable to reproduce the full set
of constraints in the stellar populations of massive ETGs -- resting
on the assumption that the analysis of gravity-sensitive line
strengths leads to a mass fraction at birth in stars with mass
$M<0.5M_\odot$ above 60\%.  Most notably, the large amount of
metal-poor gas locked in low-mass stars during the early, strong
phases of star formation results in average stellar metallicities
[M/H]$\lesssim -0.6$, well below the solar value. The conclusions are
unchanged if either the low-mass end cutoff, or the pivot mass are
left as free parameters, strengthening the case for a time-dependent
IMF.
\end{abstract}


%%%%%%%%%%%%%%%%%%%%%%%%%%%%%%%%%%%%%%%%%%%%%%%%%%%%%%%
\section{Introduction}
\label{se:intro}

The distribution of stellar masses in galaxies, i.e. the stellar
initial mass function (IMF), is of fundamental importance
for describing stellar populations. Amongst other things, the IMF
drives the chemical evolution of galaxies and defines the stellar
mass-to-light ratios (M$_\star$/L). The IMF has been intensely studied
in the Milky Way (MW) and nearby dwarf galaxies, and it has been found
mostly invariant for a large range of physical parameters
\citep{Kr02,Chab:03,KWP13}.

However, in recent years this universality of the IMF has been called
into question by a number of new observational results \citep[see,
  e.g.][]{CVC03,HG:08,MWK09,GHS10,VC10,VC12,CMA12,FBR13,WFVB:13}. On
a first glance, not all of these results agree well with each
other. While, for example, \citet{GHS10} find that the IMF becomes
increasingly top-heavy in galaxies with a high star formation rate,
other studies suggest bottom-heavy IMFs for massive early galaxies
(ETGs), systems which are believed to have formed in a massive
starburst. More hints about top-heavy IMFs have also been found in the
bulges of M31 and the MW \citep{BKM07} as well as in massive globular
clusters and ultracompact dwarf galaxies \citep{DKP12,MKD12}.

The rise of large and deep galaxy surveys in recent years opened a new
angle on stellar populations in galaxies. For example, \citet{CMA12}
used integral field spectroscopy and photometry of a volume-limited
catalogue of 260 ETGs to constrain the populations in these
galaxies. They found that, independent of the choice of dark matter
halo model for the galaxies, the SDSS $r$-band M/L ratios suggest a
systematic transition from a standard IMF \cite[e.g.][]{Kr01} towards
a distribution with heavier stellar M/L in the more massive galaxies
(by $\sim$60\% at $\sigma\sim 300$\,km\,s$^{-1}$).
Bottom-heavy IMFs for ETGs had already been suggested by studies of
abundance line indicators in ETGs \citep{CVC03} and bulges of
late-type galaxies \citep{FPV03}. The need for non-standard IMFs had
also been shown from abundance ratios and the colours of ETGs
\citep{VCP96,VPB97}. More recently, \citet{VC10,VC12} analysed the
gravity-sensitive indices Na8190 and FeH0.99 and found strong evidence
for a bottom-heavy IMF in ETGs.

This view is however not without challenge. \citet{SL13} showed a
strong lens massive ETG and compared the stellar lensing mass with
mass estimates from population synthesis modelling, They found no
striking deviation from a standard IMF. This galaxy is, however, very
extended whereas compactness has been suggested \citep{Laesker:13} as
a driver for the IMF changes.

%%%%%%%%%%%%%%%%%%%%%%%%%%%%%%%%%%%%%%%%%%%%%%%%%%%%%%%
Galactic chemical enrichment provides an additional constraint on the
IMF, as the distribution of stellar masses plays an essential role in
the enrichment of stellar populations. In \citet{WFVB:13} it was shown
that a time-independent standard bimodal IMF -- with a power-law at
the high mass end, and a smooth tapering at low masses, as defined in
\citet{VCP96} -- was not capable of explaining all the observables of
massive ETGs. The assumption of a bottom-heavy IMF results in low
overall metallicities for the stellar populations, in contradiction
with the observations. The solution, first proposed in
\citet{VCP96,VPB97}, and revisited in \citet{WFVB:13} involves a
time-dependent IMF, where a top-heavy phase, expected during the first
stages of evolution, is followed by a bottom-heavy phase. This
scenario is physically motivated by the fact that the strong
starbursts expected in massive galaxies at high redshift inject vast
amounts of energy into the ISM, changing the physical conditions to a
highly turbulent medium at high pressure, perhaps inducing a strong
fragmentation process \citep[][]{Hopkins:13,Chab:14}. However, there is an
additional aspect not covered in \citet{WFVB:13}: the chosen bimodal
IMF tightly links, by construction, the high-mass end to the low-mass
end, so one could still envision a distribution where the high-mass
end is kept at a Salpeter-like value \citep{Salp:55}, whereas the
slope at the low-mass end is left as a free parameter. Such a
functional form of the IMF is adopted by some of the groups in this
field \citep[see, e.g.][]{Cvdk:12}. In this letter, we explore the
consequences of this approach from the point of view of galactic
chemical enrichment, and we find a significant mismatch to explain
simultaneously the age-, metallicity- and gravity-sensitive features
of massive galaxies. Therefore, the need for a time-dependent IMF is
more compelling.

This letter is structured as follows: \S\ref{se:IMFfcn} provides a
generic working definition for the functional form of the IMF. In
\S\ref{se:model} the chemical evolution model with the variable IMF
used here is described. The results from our model calculations are
presented and discussed in \S\ref{se:disc}. Finally, \S\ref{se:conc}
summarizes the conclusions.  For reference, Tab.~\ref{tab:constraints}
shows conservative constraints from the literature on the observables
used to test the hypothesis of a time-independent IMF.

%%%%%%%%%%%%%%%%%%%%%%%%%%%%%%%%%%%%%%%%%%%%%%%%%%%%%%%%%%%%%%%%%
\begin{figure}
\begin{center}
\includegraphics[width=7.5cm]{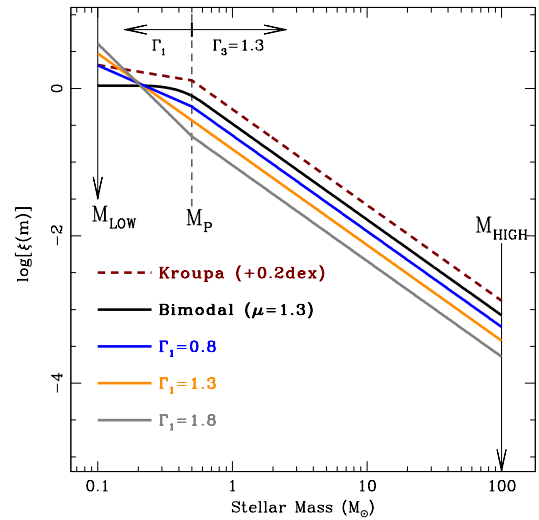}
\end{center}
\caption{Functional form of the Initial Mass Function adopted here,
  along with standard choices from the literature (see text, and
  equation 1 for details).}
\label{fig:IMF}
\end{figure}
%%%%%%%%%%%%%%%%%%%%%%%%%%%%%%%%%%%%%%%%%%%%%%%%%%%%%%%%%%%%%%%%%

%%%%%%%%%%%%%%%%%%%%%%%%%%%%%%%%%%%%%%%%%%%%%%%%%%%%%%%
\section{A functional form of the IMF}
\label{se:IMFfcn}

We assume a simple description for the initial mass function, as a 
truncated power law, namely\footnote{Note we use the $1,3$ subindex notation
as in, e.g. \citet{KWP13}. The missing subindex $2$ would refer to an intermediate
mass region that we do not consider here.} (see Fig.~\ref{fig:IMF}):
\begin{equation}
\xi(m)\equiv\frac{dN}{dm}=
\left\{
\begin{aligned}
N_1m^{-(1+\Gamma_1)}, & \qquad M_{\rm LOW}<m\leq M_{\rm P}\\
N_3m^{-(1+\Gamma_3)},    & \qquad M_{\rm P}<m<M_{\rm HIGH},\\
\end{aligned}
\right.
\end{equation}
where the slopes are given with respect to log-mass units.  We assume
the high-mass end branch to have a Salpeter-like slope,
i.e. $\Gamma_3=1.3$. Our fiducial case fixes the mass scale of the pivot
point at $M_{\rm P}=0.5M_\odot$; and the mass range: $M_{\rm
  LOW}=0.1M_\odot$; $M_{\rm HIGH}=100M_\odot$.  Note that the case
$\Gamma_1=1.3$ is similar to a \citet{Salp:55} IMF. Variations of this
expression have been used in the literature to describe the IMF
\citep[see, e.g.,][]{MillerScalo:79,Kr01,MKD12} and have been used to
explain the signature of low-mass stars in the spectra of ellipticals
\citep[e.g.][]{Cvdk:12} as well as analyses of the underlying chemical
enrichment associated to these systems
\citep{VCP96,VPB97,Bekki:13}. Spectroscopic data from
gravity-sensitive features can only constrain the mass fraction in
low-mass stars at birth \citep[see Fig.~21 of][]{FLB:13}, leaving a
significant uncertainty with respect to the stellar mass-to-light
ratio \citep{FBR13}, and complicating comparisons with
dynamically-based studies \citep[e.g.][]{CMA12,Smith:14}. The
assumption of a functional form of the IMF is needed for an adequate
comparison.

%%%%%%%%%%%%%%%%%%%%%%%%%%%%%%%%%%%%%%%%%%%%%%%%%%%%%%%%%%%%%%%%%
\begin{figure}
\begin{center}
\includegraphics[width=8.5cm]{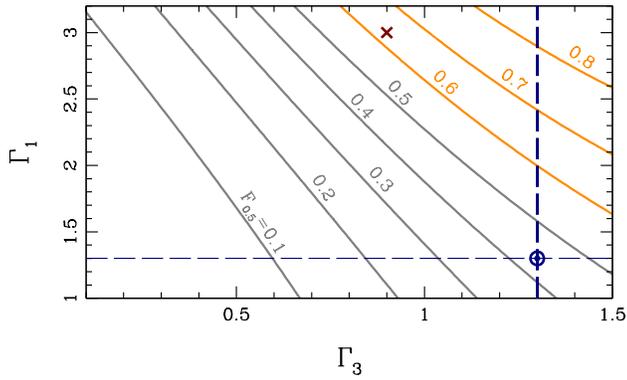}
\end{center}
\caption{Contours of $F_{0.5}$ as a function of the two IMF slopes,
  $\Gamma_1$ and $\Gamma_3$ (see text for details).  The thick
  vertical dashed line represents the fiducial set of models explored
  in this letter, along with a top- + bottom-heavy model (cross, see
  \S3).  The circle corresponds to a standard Salpeter-like
  IMF. Values $F_{0.5}\gtrsim 0.6$ are expected for massive ETGs
  (orange contours).}
\label{fig:F05}
\end{figure}
%%%%%%%%%%%%%%%%%%%%%%%%%%%%%%%%%%%%%%%%%%%%%%%%%%%%%%%%%%%%%%%%%

In \citet{FLB:13} it was found that regardless of the choice of IMF,
the gravity-sensitive spectral features imply a conservative
constraint on the stellar mass fraction, {\sl at birth}, in stars with
M$<$0.5M$_\odot$ ($F_{0.5}$). Massive ETGs are found to have
$F_{0.5}\gtrsim 0.6$.  With the assumptions adopted here, no model
with $\Gamma_1<1.5$ is capable of explaining the IMF-sensitive
spectral features of massive ETGs (see Fig.~\ref{fig:F05}).  

%%%%%%%%%%%%%%%%%%%%%%%%%%%%%%%%%%%%%%%%%%%%%%%%%%%%%%%%%%%%%%%%%
\begin{figure}
\begin{center}
\includegraphics[width=6.5cm]{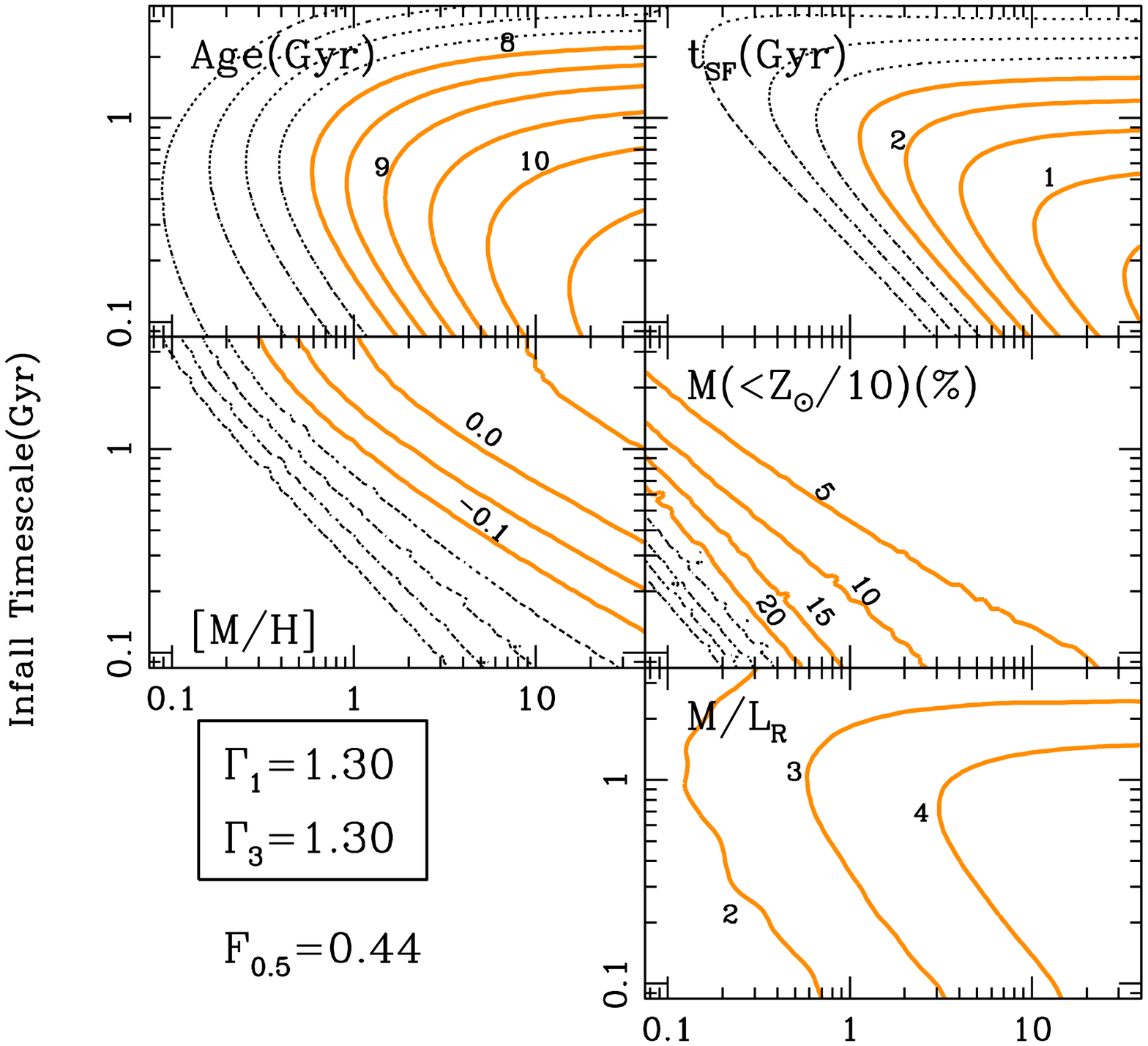}\\
\includegraphics[width=6.5cm]{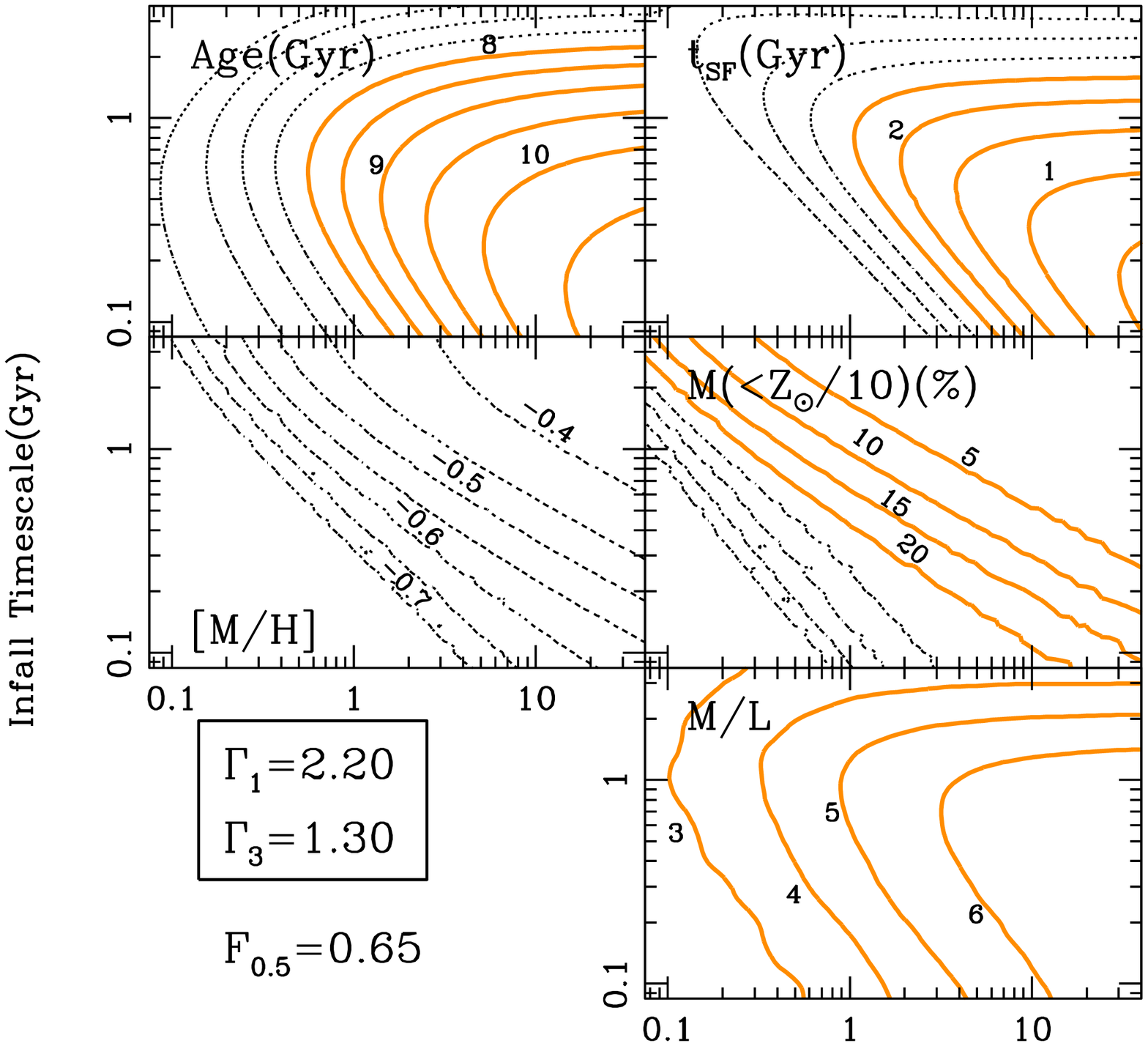}\\
\includegraphics[width=6.5cm]{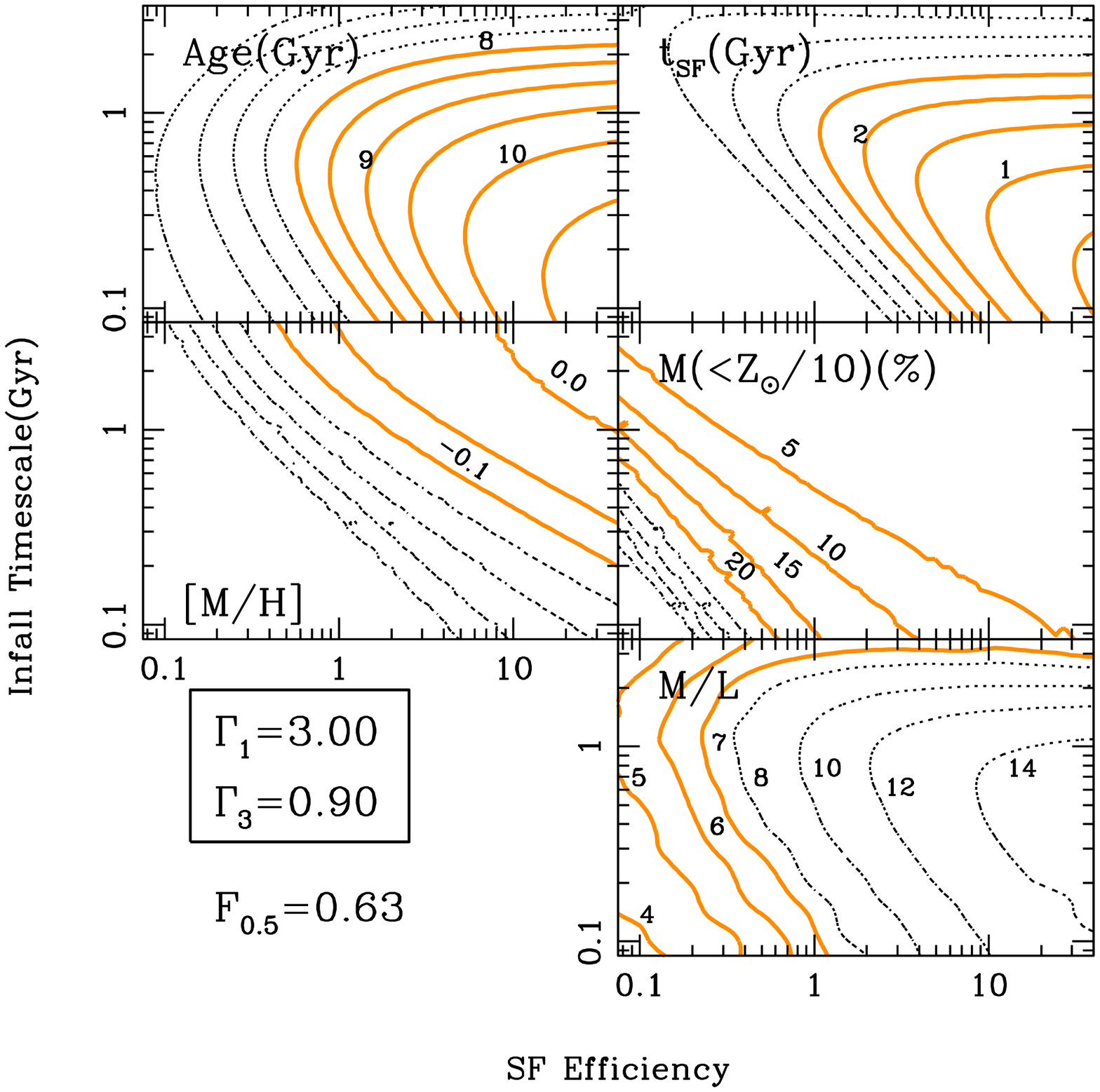}
\end {center}
\caption{Contour plots with the main observable constraints from
  the chemical enrichment code corresponding to a massive early-type
  galaxy (i.e. with a stellar mass $\gtrsim 10^{11}$M$_\odot$).  Three
  different choices of IMF slope are considered, a Salpeter-like case
  ({\sl top}), a bottom-heavy case ({\sl middle}), and a top +
  bottom-heavy case ({\sl bottom}). The contours that are compatible
  with the observational constraints (Tab.~\ref{tab:constraints}) are
  shown as thick orange lines, the remainder as dotted contours. The
  contours in age and t$_{\rm SF}$ are spaced in intervals of
  0.5\,Gyr. The contour intervals for $[{\rm M/H}]$ and
  M$(<Z_\odot/10)$ are 0.05\,dex and 5\%, respectively. All the
  contours in M/L are labelled, and measured with respect to the
  SDSS-r band. See text, and Tab.~\ref{tab:constraints} for details.} 
\label{fig:grids}
\end{figure}
%%%%%%%%%%%%%%%%%%%%%%%%%%%%%%%%%%%%%%%%%%%%%%%%%%%%%%%%%%%%%%%%%

%%%%%%%%%%%%%%%%%%%%%%%%%%%%%%%%%%%%%%%%%%%%%%%%%%%%%%%
\section{Chemical enrichment modelling and observational constraints}
\label{se:model}

We explore a set of phenomenological models tracking galactic
chemical enrichment through a reduced set of parameters. These models
are presented in detail in \citet{FS:00a,FS:00b}. In
\citet{WFVB:13} we apply them to a time-dependent IMF in order to
explain the gravity-sensitive line strengths found in massive ETGs.
In a nutshell, the buildup of the stellar component of a galaxy is
described by four parameters: a gas infall timescale ($\tau_f$), a
star formation efficiency ($C_{\rm eff}$), that follows a Schmidt law,
a formation redshift ($z_{\rm FOR}$) at which the whole process
starts, and a fraction of gas ejected in outflows ($B_{\rm out}$). 

We ran a grid of models adjusted to the stellar populations in massive
ETGs \citep[see, e.g.][]{IGDR:11}. In order to achieve a homogeneously
old population, we need to assume an early start for the star
formation process ($z_{\rm FOR}=3$), and negligible outflows ($B_{\rm
  out}=0$). Changes in these two parameters will mostly induce an
overall shift in the average age and metallicity, respectively.
Furthermore, non-neglible outflows would produce {\sl lower}
metallicities, hence strengthening our conclusion towards a
time-dependent IMF (see \S4). The other two parameters, namely the gas
infall timescale ($\tau_f$) and star formation efficiency ($C_{\rm
  eff}$) are left as free parameters in the grid. The model grids are
run for a range of IMF slopes, $1<\Gamma_1<2.5$.

Fig.~\ref{fig:grids} shows the results of the chemical enrichment
modelling for three choices of IMF slope: Salpeter
($\Gamma_1=\Gamma_3=1.3$, left); bottom-heavy ($\Gamma_1=2.2$,
$\Gamma_3=1.3$, middle), and the additional case of a top + bottom
heavy IMF ($\Gamma_1=3$, $\Gamma_3=0.9$, right).  For the latter, we
modify the high-mass slope, $\Gamma_3$, to values that would be
compatible with the observations of top-heavy IMFs in star-forming
systems \citep[e.g.][]{GHS10}, changing in addition the low-mass
slope, $\Gamma_1$, to accommodate the high values of $F_{0.5}$.  For
each case, we show (counter-clockwise from the bottom-right) contours
of stellar mass-to-light ratio (in the SDSS-$r$ band); mass fraction
in low-metallicity stars (M$(<Z_\odot/10)$); star formation lapse
(t$_{\rm SF}$); average age, and average metallicity ([M/H]). Averages
are mass-weighted.  The star formation lapse is defined as the time
period between the 25th and 75th percentiles of the stellar mass
build-up in the galaxy.  t$_{\rm SF}$ is therefore tightly linked to
[Mg/Fe]. As a very conservative estimate, we assume that values
$t_{\rm SF}\gtrsim 2$\,Gyr are in conflict with the observed
over-abundance of [Mg/Fe] in massive ETGs \citep{IGDR:11}.
M$(<Z_\odot/10)$ is defined as the fraction of stellar mass with
metallicity below 1/10 of solar. This is an indicator of the G-dwarf
problem \citep{Gdwarf}. It is a well-known fact that some models can
produce old, and overall metal-rich populations, although with a
significant tail of low-metallicity stars. Models with a mass fraction
over 10-20\% in low-metallicity stars would be in conflict with the
observations of massive ETGs
\citep{VCP96,VPB97,Maraston:00,Nolan:07}.

%%%%%%%%%%%%%%%%%%%%%%%%%%%%%%%%%%%%%%
%%%%%%%%%%   TABLE 1   %%%%%%%%%%%%%%%
%%%%%%%%%%%%%%%%%%%%%%%%%%%%%%%%%%%%%%
\begin{table}
\begin{center}
\begin{tabular}{ccc}
\hline
Observable & Constraint & Reference\\
\hline
 Age (Gyr)         & $[8,10]$ & (1)\\
$[{\rm M/H}]$      & $[-0.1,+0.2]$  & (1)\\
 t$_{\rm SF}$ (Gyr) & $[0.5,2.0]$ & (2)\\
 M$_s(<Z_\odot/10)$ & $[0.05,0.20]$ & (3)\\
 F$_{0.5}$          & $[0.6,0.8]$ & (4)\\
$\Upsilon_r/\Upsilon_{r,\odot}$    & $<7.0$ & (4)\\
\hline 
\end{tabular}\\
\end{center}
\caption{Constraints on the general properties of the unresolved
  stellar populations in massive early-type galaxies. The
  uncertainties are rough estimates, quoted at a conservative
  1$\sigma$ level. The references are: (1) \citet{Trager:00,Thomas:05}; 
  (2) \citet{IGDR:11}; (3) \citet{VPB97}; (4) \citet{FLB:13} }
\label{tab:constraints}
\end{table}
%%%%%%%%%%%%%%%%%%%%%%%%%%%%%%%%%%%%%%%%%%%%%%%%%%%%%%%%%%%

%%%%%%%%%%%%%%%%%%%%%%%%%%%%%%%%%%%%%%%%%%%%%%%%%%%%%%%
\section{Results \& Discussion}
\label{se:disc}

Fig.~\ref{fig:grids} shows that the Salpeter model ({\sl left}) is
capable of recreating the old, metal rich, [Mg/Fe] over-abundant
populations without a significant low-metallicity trail. However, the
fraction in low-mass stars ($F_{0.5}=0.44$) is in conflict with the
recent interpretation of gravity-sensitive spectral features
\citep{FLB:13}. The top- + bottom-heavy model ({\sl right}) is
challenged by the overly high values of M/L in the region of parameter
space compatible with the age/metallicity constraints.  The same
result would hold if we chose to keep $\Gamma_3$ as a free
parameter. Therefore, we rule out the option of a change in both
slopes, and hereafter focus on the case where the high-mass end slope
is fixed to the Salpeter-like value ($\Gamma_3=1.3$).  The
bottom-heavy model ({\sl middle}) allows for a higher low-mass
fraction ($F_{0.5}=0.65$) at the price of locking too much gas in
low-mass stars during the early (metal-poor) stages. The average
metallicities are significantly lower than those derived from the
observations in massive ETGs.

\begin{figure}
\begin{center}
\includegraphics[width=8.cm]{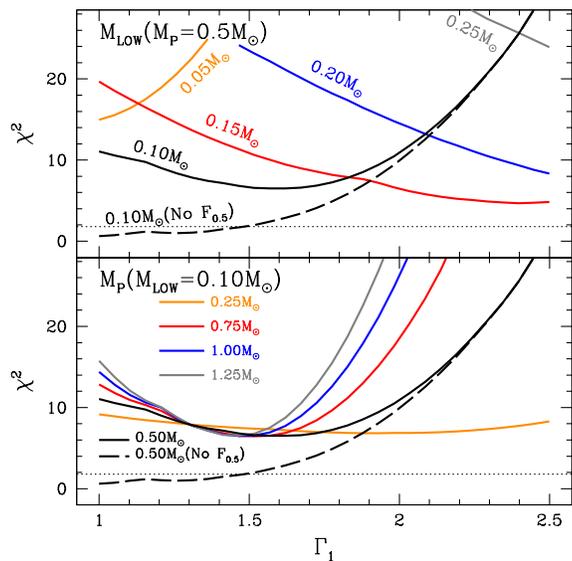}
\end{center}
\caption{Fits to the extended models, where the low-mass end of the
  IMF ({\sl top}), or the pivot mass ({\sl bottom}) are considered as
  free parameters in a time-invariant IMF. The $\chi^2$ values are
  shown for several cases (solid lines). We illustrate the
  constraining power of $F_{0.5}$ by including the case where this
  constraint is removed from the analysis (dashed lines).  For
  reference, the $\chi^2$ of a time-dependent IMF \citep[model~A
    in][]{WFVB:13} is shown as a dotted horizontal line in both panels
  (where all constraints from Tab.~\ref{tab:constraints}, including
  F$_{0.5}$, are imposed).  }
\label{fig:chi2}
\end{figure}

One could consider additional changes of the functional form of the
IMF, most notably changing either the pivot point ($M_{\rm P}$) --
motivated by a change in the physical properties of the star forming
regions \citep[see, e.g.][]{Larson:05} -- or the low-mass end ($M_{\rm
  LOW}$) -- as suggested in the recent analysis of ETG lenses
\citep{Barnabe:13}. Note that we emphasize in this letter that all
these changes would relate to an otherwise {\sl time-independent}
IMF. However, from the previous figure, it is expected that such
changes nevertheless lock large masses of low-metallicity stars,
leading to results that are incompatible with the metal-rich
populations found in massive galaxies. To further illustrate this
point, we show in Fig~\ref{fig:chi2} a $\chi^2$ estimator based
on the observational constraints presented above (see
Tab.~\ref{tab:constraints}), where Gaussian constraints are imposed at
the 1\,$\sigma$ level over the allowed intervals, in effect producing
very conservative limits. For M/L -- where the constraint derives from
{\sl dynamical} M/L measurements -- we simply penalize the likelihood
for values higher than $\Upsilon_r=7\Upsilon_{r,\odot}$ using a Gaussian with
$\sigma=0.5$. We consider several cases, where either a
range for the low-mass end (M$_{\rm LOW}$, top panel) or the pivot
mass (M$_{\rm P}$, bottom panel) are explored in a time-independent
IMF. The solid lines correspond to different cases where all four
constraints are imposed.  The dashed lines show, for comparison, the
case where the $F_{0.5}$ constraint is removed from the analysis. As
reference, the horizontal dotted line in both panels represent the
fiducial, time-dependent model\footnote{We emphasize that the
  functional form of the (bimodal) IMF in \citet{WFVB:13} changes both
  low- and high-mass ends with a single parameter, $\Gamma$.}~A of
\citet{WFVB:13}, where a top-heavy IMF is followed by a sharp
transition towards a bottom-heavy IMF after 0.3\,Gyr. Note the
different behaviour with respect to changes of either the low-mass end
of the IMF, or the pivot mass scale. In the top panel, a transition is
apparent at M$_{\rm LOW}=0.10$M$_\odot$, above which very bottom-heavy
IMFs seem to be favoured, with the best case at
$0.15$M$_\odot$. Nevertheless, even this option is rejected with
respect to a time-dependent scenario.

Fig.~\ref{fig:chi2} confirms that for a wide range of options, a
time-independent IMF is incompatible with constraints from galactic
chemical enrichment.  Although the proposal presented in
\citet{WFVB:13} is a simple toy model, this letter supports the need
for a time-dependent mechanism that tips an initial top-heavy IMF in a
strongly star-forming system, towards a bottom-heavy IMF during the
final stages of the starburst. Such a mechanism would reconcile the
apparent contradiction between the observations of star-bursting
systems and the properties of quiescent galaxies that underwent a
star-bursting phase in the past. We emphasize that the relatively long
duration ($\sim 1-2$\,Gyr) of a strong star formation phase in massive
ETGs at high redshift is expected to cause this transition.

%%%%%%%%%%%%%%%%%%%%%%%%%%%%%%%%%%%%%%%%%%%%%%%%%%%%%%%%%%%%%%%%%
%%%%%%%%%%%%%%%%%%%%%%%%%%%%%%%%%%%%%%%%%%%%%%%%%%%%%%%%%%%%%%%%%

%%%%%%%%%%%%%%%%%%%%%%%%%%%%%%%%%%%%%%%%%%%%%%%%%%%%%%%
\section{Conclusions}
\label{se:conc}

We explore a model of galactic chemical enrichment, with the
assumption of a time-independent IMF, with several free parameters
controlling the contribution of low- and high-mass stars. The
functional form (presented in \S\ref{se:IMFfcn}) is representative of
the typical functions explored in the literature. A comparison of our
models with conservative observational constraints in massive ETGs
(Tab.~\ref{tab:constraints}) reject this hypothesis, mainly based on
the locking of too many low-mass metal-poor stars during the first
phases of formation, and on the assumption that the recent
observations of gravity-sensitive line strengths result in a
constraint on the mass fraction in low-mass stars at birth
(F$_{0.5}>0.6$). The best-fit time-independent models give stellar
metallicities [M/H]$\sim -0.6$, i.e. significantly lower than the
observational constraints -- although higher than the metallicites
expected for a time-independent ``bimodal'' IMF as defined in
\citet{VCP96}.  In contrast, a simple time-dependent model, justified
by the large energy injection during the strong star bursting phase in
massive galaxies, leads to a consistent picture with an overall old,
metal-rich and bottom-heavy stellar population, as suggested by
\citet{VCP96,VPB97} and, more recently by \citet{WFVB:13}.

%%%%%%%%%%%%%%%%%%%%%%%%%%%%%%%%%%%%%%%%%%%%%%%%%%%%%%%

\section*{Acknowledgements}
We would like to thank the referee, Dr Russell Smith for very valuable
comments and suggestions about this letter. CW and AV acknowledge
support from grant AYA2013-48226-C3-1-P from the Spanish Ministry of
Economy and Competitiveness (MINECO).

%%%%%%%%%%%%%%%%%%%%%%%%%%%%%%%%%%%%%%%%%%%%%%%%%%%%%%%
\bibliography{mybiblio}

\bsp
\label{lastpage}
\end{document}